\documentstyle[11pt,aaspp4]{article}
\newcommand\beq{\begin{equation}}
\newcommand\eeq{\end{equation}}

\begin{document}

\title{Disks irradiated by beamed radiation from compact objects}

\author{Rosalba Perna\altaffilmark{1} and Lars Hernquist }
\medskip
\affil{Harvard-Smithsonian Center for Astrophysics, 60 Garden Street,
Cambridge, MA 02138}

\altaffiltext{1}{Harvard Junior Fellow}

\begin{abstract}
We examine the reprocessing of X-ray radiation from compact objects by
accretion disks when the X-ray emission from the star is highly
beamed.  The reprocessed flux for various degrees of beaming and
inclinations of the beam axis with respect to the disk is
determined. We find that, in the case where the beam is produced by a
non-relativistic object, the intensity of the emitted spectrum is
highly suppressed if the beam is pointing away from the disk. However,
for beams produced by compact objects, general relativistic effects
cause only a small reduction in the reradiated flux even for very
narrow beams oriented perpendicularly to the disk. This is especially
relevant in constraining models for the anomalous $X$-ray pulsars,
whose $X$-ray emission is highly beamed.  We further discuss other
factors that can influence the emission from disks around neutron
stars.
\end{abstract}


\section{Introduction}

Reprocessing of $X$ rays by accretion disks is a common phenomenon in
astrophysics.  For example, reprocessed radiation is thought to be
responsible for the optical emission observed in $X$-ray binaries
(e.g. van Paradijs \& McClintock 1994; de Jong, van Paradijs \&
Augusteijn 1996) and the ratio between optical and $X$-ray flux has
been connected with properties of the system itself, such as the
orbital period (van Paradijs \& McClintock 1994).

Another class of astrophysical objects for which reprocessing of
radiation by an accretion disk could be potentially important is that
of the Anomalous $X$-ray Pulsars (AXPs). The luminosity of these
objects, $L_x\sim 10^{35-36}$ erg ${\rm s}^{-1}$, is far larger than their rotational
energy loss, $|\dot{E}|\equiv 4\pi^2 I \dot{P}/P^3 \approx 10^{32.5}$
erg ${\rm s}^{-1}$, and therefore their X-ray
emission cannot be rotation-powered.
Competing models to explain the origin of their $X$-ray luminosity
invoke either the existence of neutron stars (NSs) with very large
magnetic fields, $B\sim 10^{14-15}$ G (e.g. Duncan \& Thompson 1992;
Thompson \& Duncan 1996;
Heyl \& Hernquist 1998a,b) or accretion from either a companion
(Mereghetti \& Stella 1995), or a disk; where the latter situation
could have
resulted from
the debris of a disrupted high mass companion (van Paradijs, Taam \&
van den Heuvel 1995; Ghosh, Angelini \& White 1997) or from material
falling back after the supernova explosion (Corbet et al. 1995;
Chatterjee, Hernquist \& Narayan 2000; Alpar 1999, 2000; Marsden et
al. 1999).  So far, no evidence for a companion has been found
for any AXP, and
the debate on whether the AXPs are isolated neutron stars powered by
internal energy or NSs accreting from a disk of some sort is still
open.

The strongest constraint on the presence of a disk is given by its
optical and longer wavelength emission which, at the luminosities
typical for AXPs, is dominated by reradiation.  Phase observations of
the $X$-ray luminosity show a high degree of pulsation, which cannot
be accounted for by an isotropic $X$-ray emission, but require some
degree of beaming (DeDeo, Psaltis \& Narayan 2000; Perna, Heyl \&
Hernquist 2000). However, computations of the reprocessed flux for
disks around AXPs (Perna, Hernquist \& Narayan 2000; Hulleman et
al. 2000) have so far always assumed isotropic emission from the
star. In this {\em Letter}, we consider the problem of disks
irradiated by beamed radiation. For radiation produced by a compact
object, such as a neutron star, the general relativistic effect of light
bending is taken into account in transforming the beam pattern
formed within the atmosphere of the neutron star into that which is
seen by an observer at infinity. We show that the inclusion of these
relativistic corrections has a significant influence on the
reradiated spectrum when the beam points away from the disk. We
finally discuss, more generally, other effects that can influence the
spectrum of the reradiated flux, particularly in the context of the
types of disks expected in accretion models for AXPs.

\section{Model}

\subsection{Beaming pattern}

The radiation pattern emerging from accretion into the atmosphere of a 
neutron star (NS) has been computed in detail
by Nagel (1981) and Meszaros \& Nagel (1985). At low accretion rates, a good
approximation is given by the function 
\beq
I(\delta)=I_0 \cos^n(\delta)\;,
\label{eq:Id}
\eeq
with $n\simeq 2-3$,
where $\delta$ is the angle that the emitted photons make with the
normal to the local stellar surface. 
What we need to find is the pattern of radiation that an
observer at infinity will detect. Due to light bending effects, a photon 
emitted at an angle $\delta$ with respect to
the normal to the surface comes from a 
colatitude $\theta$ on the star given by (e.g. Page 1995)
\beq
\theta(\delta)=\int_0^{R_s/2R}x\;du\left/\sqrt{\left(1-\frac{R_s}{R}\right)
\left(\frac{R_s}{2R}\right)^2-(1-2u)u^2 x^2}\right.\;,
\label{eq:teta}
\eeq
where $x\equiv\sin\delta$, $R$ is the radius of the star, and
$R_s=2GM/c^2$ its Schwarzchild radius.  If the beam pattern formed in
the atmosphere of the NS is $I(\delta)$, then an observer at infinity
will see a beam pattern given by
\beq
I(\theta)=I(\delta)\frac{\sin\delta}{\sin\theta}\frac{d\delta}{d\theta}\;.
\label{eq:It}
\eeq
Given a colatitude $\theta_*$ on the star, the corresponding angle $\delta_*$
at the emission point is computed by numerically solving the equation 
$\theta_*-\theta(\delta)=0$. The intensity at that point is then found from
Equ.~({\ref{eq:Id}), and converted to the observed radiation emitted at
$\theta_*$ with the transformation given in Equ.~(\ref{eq:It}). 

Figure 1 shows $I(\theta)$ for a star of mass $M=1.4 M_\odot$ and different
choices of radii. The more relativistic a star is (i.e. smaller $R/R_s$),
the more spread out the radiation pattern becomes. 

Now, let $\alpha$ be the angle that the axis of the beam makes with the plane of the
disk; the axis is then described by the unit vector $\hat{O}=
(\cos\alpha\cos\phi,\cos\alpha\sin\phi,\sin\alpha)$, where $\phi=\Omega t$,
with $\Omega$ being the rotation rate of the star.  Then, let
$\hat{P}=(\sin\beta,0,\cos\beta)$ be a point along  a direction
making  an angle $\beta$ with the perpendicular to the disk in the upper plane.
While the star rotates, the angle formed between the beam axis and the
direction along  $\hat{P}$ is  
\beq
\theta_{\rm up}(\phi,\beta;\alpha)
=\arccos(\sin\beta\cos\alpha\cos\phi + \cos\beta\sin\alpha)\;.
\label{eq:teta1}
\eeq
Due to the symmetry of the problem, two beams are expected from the NS, 
centered around the two poles. If the axis of the upper beam forms an angle $\alpha$
with the plane of the disk, the axis of the lower beam is at an angle
$-\alpha$, and, while the star rotates,  it  makes an angle 
\beq
\theta_{\rm down}(\phi,\beta;\alpha)
=\arccos(\sin\beta\cos\alpha\cos\phi - \cos\beta\sin\alpha)\;.
\label{eq:teta2}
\eeq
with the direction $\hat{P}$ defined above. 

Figure 2 shows the intensity profiles averaged over
a rotation period of the star, as a function of the angle
$\beta$ of the observation point with respect to the normal to the disk,
\beq
\left<I(\beta;\alpha)\right>=\frac{1}{2\pi}\int_0^{2\pi}d\phi\;\left\{ 
I[\theta_{\rm up}(\phi,\beta;\alpha)] +I[\theta_{\rm down}(\phi,\beta;\alpha)]
\right\}\;.
\label{eq:Ib}
\eeq  

\subsection{Temperature profile and emitted spectrum}

Following Vrtilek et al. (1990), the computation of the temperature profile
is obtained by combining two equations for the disk structure.
The equation for hydrostatic equilibrium in the $z$-direction is
\beq
\frac{k}{\mu m_p} T =\omega^2h^2\;,
\label{eq:1}
\eeq
where $\omega$ is the Keplerian rotation rate, $\omega=\sqrt{GM/r^3}$,
$h$ is the half-thickness of the disk, $k$ the Boltzmann constant, 
$\mu$ the mean molecular weight, and $m_p$ the proton mass. 
The second equation for the disk structure states that the radiation
emitted by a surface element, $\sigma T_{\rm eff}^4$, should be
equal to the sum of the energy released locally and the absorbed radiation
(Cunningham 1976; Pacharintanakul \& Katz 1980)
\beq
\sigma T_{\rm eff}^4=\sigma T_0^4 + \frac{f_1 L_x}{4\pi r}\frac{\partial h/r}
{\partial r}\;,
\label{eq:2}
\eeq
where 
\beq
\sigma T_0^4=\frac{3GM\dot{M}}{8\pi r^3}\left(1-\sqrt{\frac{R}{r}}\right)
\label{eq:3}
\eeq
gives the effective temperature for a disk which is not irradiated
(Shakura \& Sunyaev 1973). In Equation (\ref{eq:2}), $f_1$ is the
absorbed fraction of the radiation impinging on the surface (the quantity $1-f_1$
is called the ``albedo'' of the disk), for which 
we assume $f_1=0.5$ as in Vrtilek at al. unless otherwise stated. 
The $X$-ray luminosity is given by 
\beq
L_x(\beta;\alpha)=L_x^0 \;\left<I(\beta;\alpha)\right>\;,
\label{eq:4}
\eeq
with $\left<I(\beta;\alpha)\right>$ given by Equation (\ref{eq:Ib}), and normalized
so that $\int_0^{\pi/2} \left<I(\beta;\alpha)\right>\sin\beta \;d\beta =1$. 
$L_x^0$ is the luminosity corresponding to isotropic emission, and it can be related to
$\dot{M}$ via
\beq
L_x=f_2 \frac{GM{\dot M}}{R}\;,
\label{eq:5}
\eeq
where a value of $f_2= 0.5$ corresponds to the assumption that the
$X$-ray luminosity produced in the disk is  radiated mainly
in a direction perpendicular to the orbital plane, and that only
the $X$-rays directly from the NS can impinge on the disk.
Finally, the angle $\beta$ in Equation (\ref{eq:4}) is related to the
half-thickness of the disk at position $r$ by
\beq
\beta = \frac{\pi}{2}-\arctan\left[\frac{h}{r}\right]\;.
\label{eq:6}
\eeq
Note that the vertical structure of the disk is assumed to be isothermal.

Equations (\ref{eq:1}) - (\ref{eq:6}) are solved numerically for the
temperature profile $T(r)$ , and the flux to the observer 
is obtained by integrating the local emissivity $B_\nu(T)$ 
(assumed black body) over the entire surface of the disk:
\beq
F=\frac{1}{d^2}\int_{r{\rm min}}^{r{\rm max}}dr\,r \int_0^{2\pi}d\phi\;
B_\nu(T) \times
[\cos\phi \sin\gamma(r) \sin i + \cos\gamma(r) \cos i]\;.
\label{eq:flux}
\eeq
Here $i$ is the angle between the line of sight and the normal 
to the disk at $r=0$, and $\gamma(r)=dh(r)/dr$ is the tilt angle of the
surface of the disk at position $r$. In our calculations we assume
 $i=60^\circ$.
The inner edge of the disk, $r_{\rm min}$, is
taken at the magnetospheric radius $R_m$: for a typical luminosity 
$L_x\sim 10^{35}$ ergs s$^{-1}$ and a magnetic field $B\sim 8\times 10^{12} G$,
as required by the accretion model of Chatterjee et al., 
one has $R_{m}\sim 2\times 10^9$ cm.   
The outer edge, $R_{\rm out}$, is determined
according to the solution given by Cannizzo (1993); for a disk of
about $\sim 10^4$ years, as  typical for the AXPs, it is\footnote{We
also assume that the disk has a high opacity, as it is likely to be
made of heavy elements if formed from the debris of a supernova explosion.} 
$R_{\rm out}\sim 
5\times 10^{14}$ cm. We show our results for an $X$-ray luminosity
$L_x=10^{35}$ ergs s$^{-1}$.

Figure 3 shows the emitted flux for the exponents $n=2$ and $n=3$  
of the beaming function, and for the limiting inclination angles 
of the beam axis $\alpha=0^\circ$ and $\alpha=90^\circ$.
Each case is made for $R/R_s\rightarrow\infty$ (non-relativistic case),
and for $R/R_s=3$.  
The emission corresponding to the isotropic case is shown for
comparison. The figure illustrates how, in the non-relativistic case, 
the reprocessed flux is suppressed by several orders of magnitude
if the beam axis is pointing away from the disk.  However, in the
case of radiation produced by a highly relativistic object such as
a neutron star, even a beam perpendicular to the disk
($\alpha=90^\circ$) gives a significant contribution to the spectrum. 
This reduction in the difference between spectra produced by beams
pointing parallel and perpendicularly to the disk is due to the same
general relativistic effects that cause a reduction in the pulsation
amplitude of the radiation from the neutron star itself (Page 1995).

\section{Discussion}

We have considered the effects that beaming of the $X$-ray luminosity
impinging on a disk has on the flux that is reradiated by the disk
itself.  We have found that, due to general relativistic effects of
light deflection, even narrow beams pointing away from the disk still
give a significant contribution to the impinging flux and on the
consequent intensity of the reprocessed radiation. 

The problem we have considered is particularly important for the AXPs,
which, due to the high pulsation amplitude of their flux, must produce
a highly beamed $X$-ray luminosity. The reradiated flux gives the
dominant contribution at long wavelengths; this is produced in the
bulk of the disk, whereas the optical emission is generated in its
innermost part.  If the inner edge of the disk were truncated at a
radius larger than the magnetospheric radius (which could for example
happen if the inner region of the disk made a transition to a
different state), then the optical emission would be highly
suppressed. This is shown in panel (a) of Figure 4, where the spectrum
is computed for different choices of $r_{\rm min}$. Constraints from
optical observations have already been made in a few cases. Hulleman et al.
have used $I$ and $R$ band observations to rule out accretion scenarios for 
the AXP 1E2259. However, if this object had an accretion disk truncated at 
$r_{\rm min}\sim 10 R_m$, the observational constraints would not be violated
any longer\footnote{Note that these constraints are given assuming
an inclination angle of $60^\circ$ between the line of sight and the
normal to the disk. Clearly, a higher inclination would not  violate
the observational constraints while allowing a larger $r_{\rm min}$.}. 

On the other hand, we have shown that the longer wavelength emission is
hardly suppressed, even if the $X$-ray emission is highly beamed and the
beam axis points away from the disk.
A reduction of the disk to 10\% of its
size (panel (b) in Figure 4), or an increase in  the disk
albedo from 50\% to 95\% (panel (c) of
Figure 4) would not reduce the peak flux by more than an order of
magnitude. Therefore, definite constraints on the nature of the
AXPs can only be made from multiwavelength observations.

\acknowledgements 
We are indebted to Ramesh Narayan and Jeremy Heyl for stimulating discussions.

\begin{figure}[t]
\centerline{\epsfysize=5.7in\epsffile{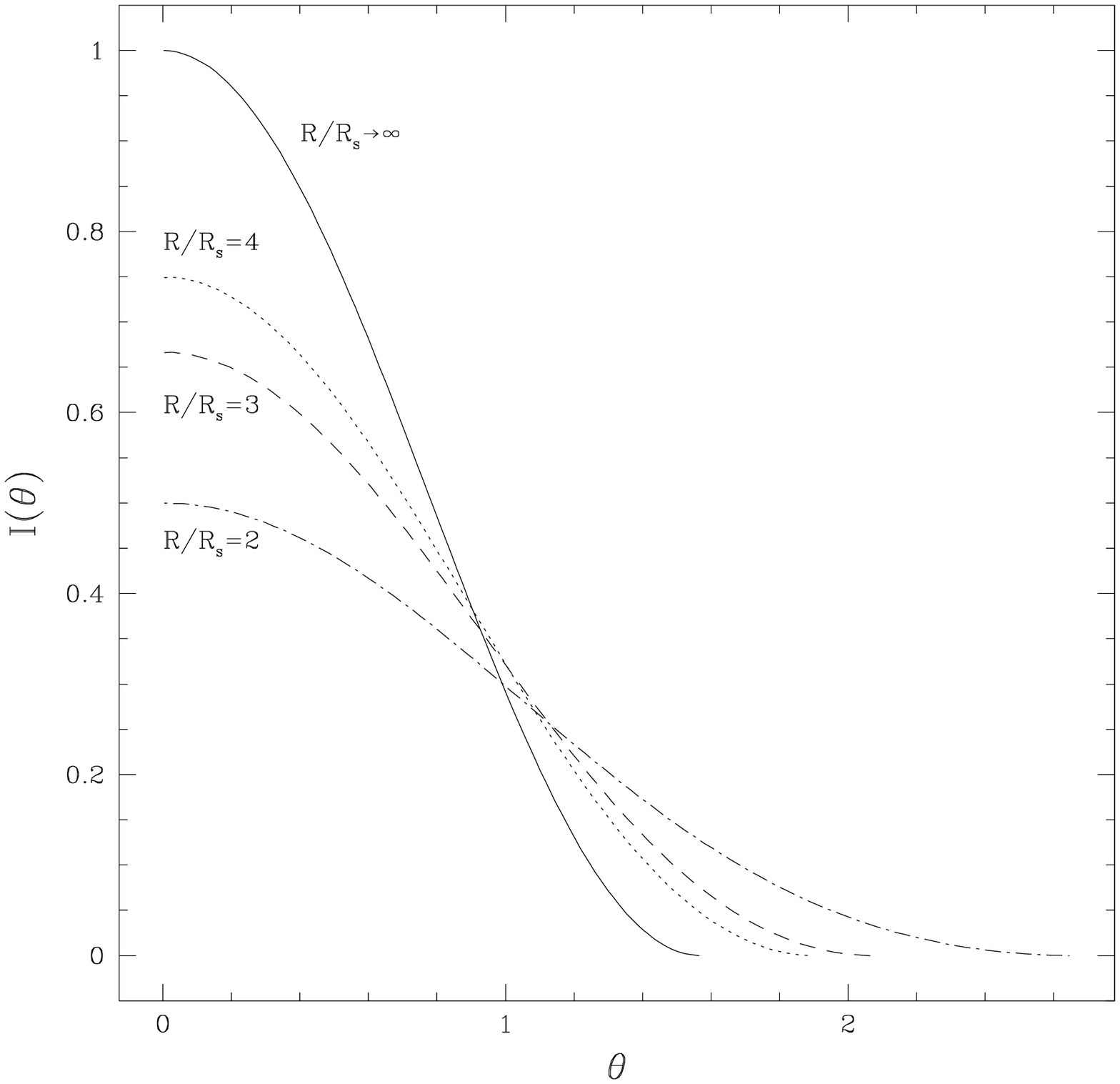}}
\caption{Beaming pattern as it would appear to an observer at infinity
for a NS of mass $M=1.4M_\odot$ and different values of its radius.
The emitted pattern at the NS surface is $I(\delta)=\cos^2\delta$.
Light deflection effects make this pattern much broader. }
\label{fig:1}
\end{figure}

\begin{figure}[t]
\centerline{\epsfysize=5.7in\epsffile{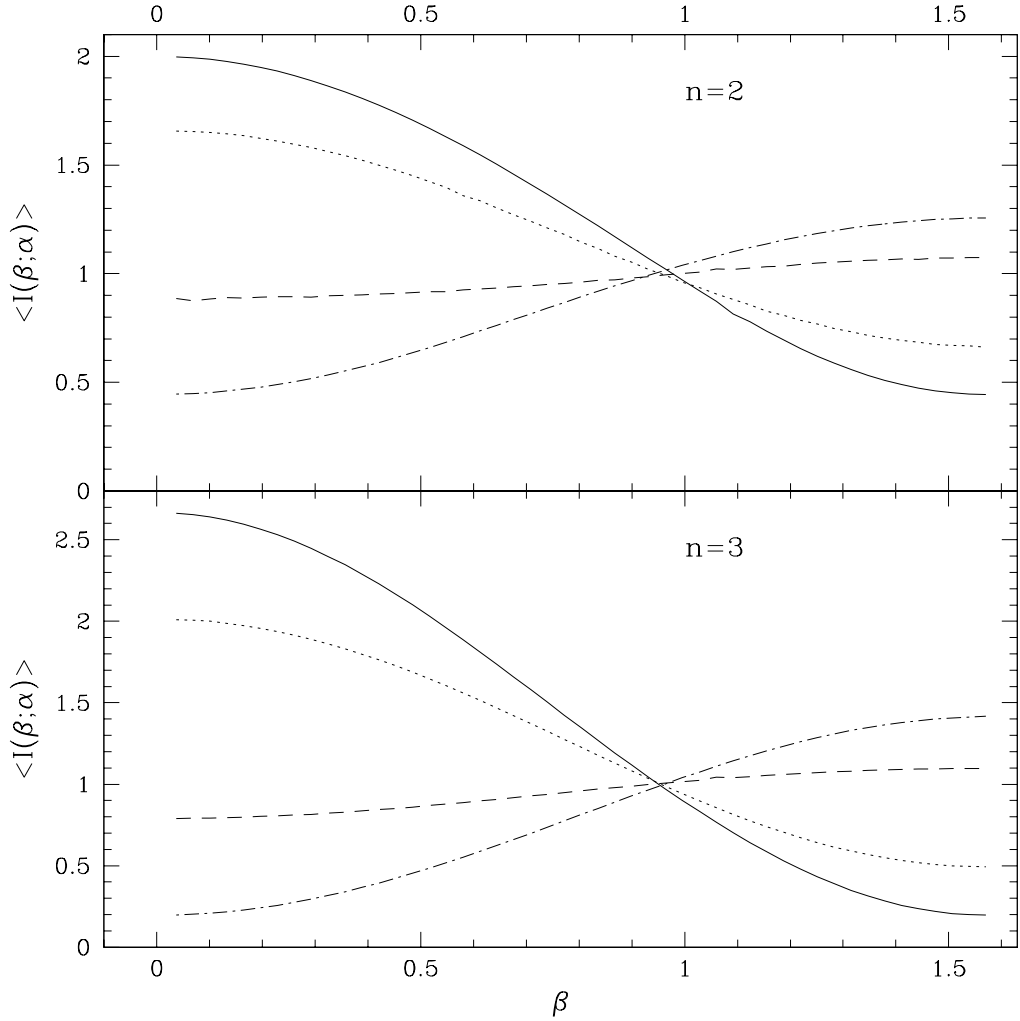}}
\caption{Average intensity over a period of the star as a function of the 
angle with the normal to the disk. Here $R/R_s=3$, and in both panels
the various curves correspond to different choices of the inclination
angle $\alpha$ of the beam axis with the plane of the disk:
$\alpha=90^\circ$ (solid line), $\alpha=60^\circ$ (dotted line), $\alpha=30^\circ$
(dashed line), $\alpha=0^\circ$ (dotted-dashed line). The intensity is normalized
to the value of isotropic luminosity.}
\label{fig:2}
\end{figure}

\begin{figure}[t]
\centerline{\epsfysize=5.7in\epsffile{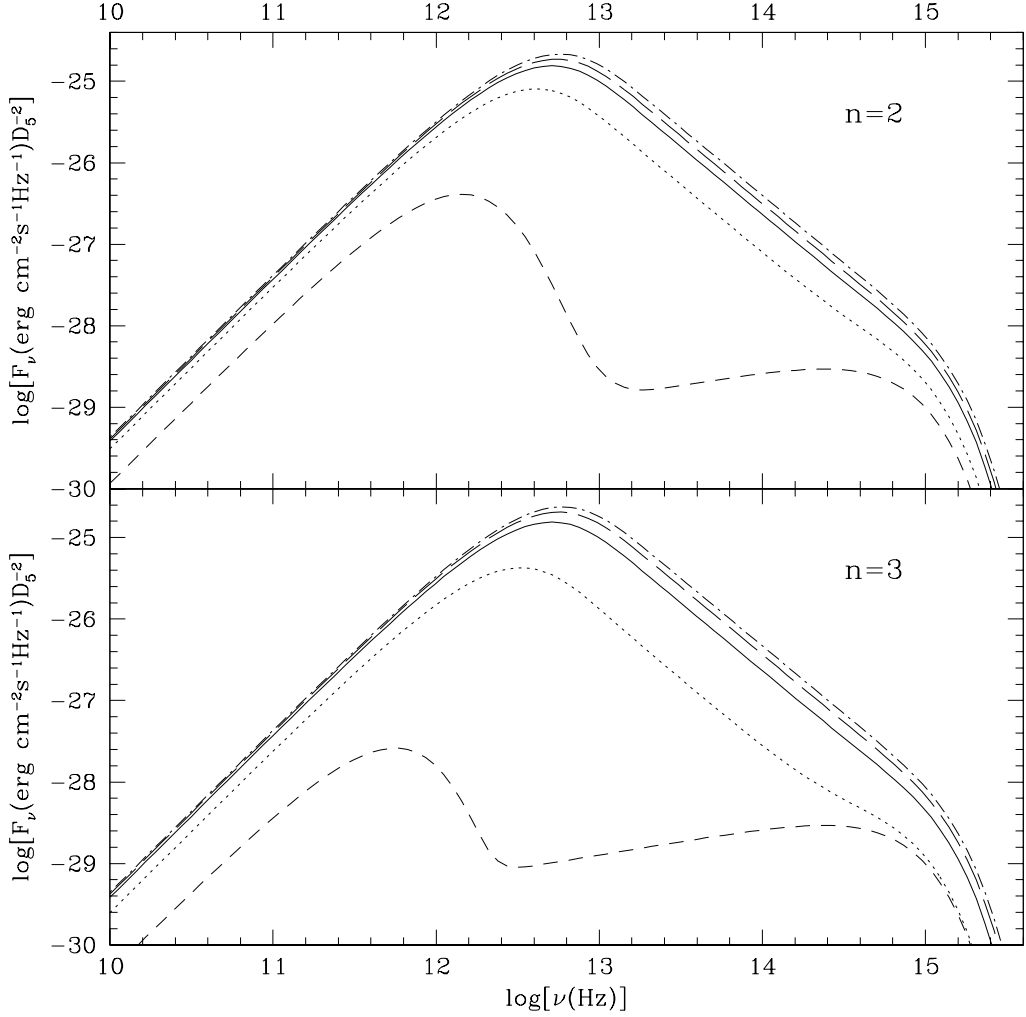}}
\caption{Emitted spectrum from a disk of age about $10^4$ yr, illuminated
by an $X$-ray luminosity $L^0_x=10^{35}$ ergs sec$^{-1}$.
In both panels, the various curves correspond to the following cases:
isotropic luminosity (solid line), $R/R_s=3$ and $\alpha=90^\circ$ (dotted line),
$R/R_s\rightarrow\infty$ and $\alpha=90^\circ$ (dashed line),
$R/R_s=3$ and $\alpha=0^\circ$ (long-dashed line), 
$R/R_s\rightarrow\infty$ and $\alpha=0^\circ$ (dotted-dashed line).}
\label{fig:3}
\end{figure}

\begin{figure}[t]
\centerline{\epsfysize=5.7in\epsffile{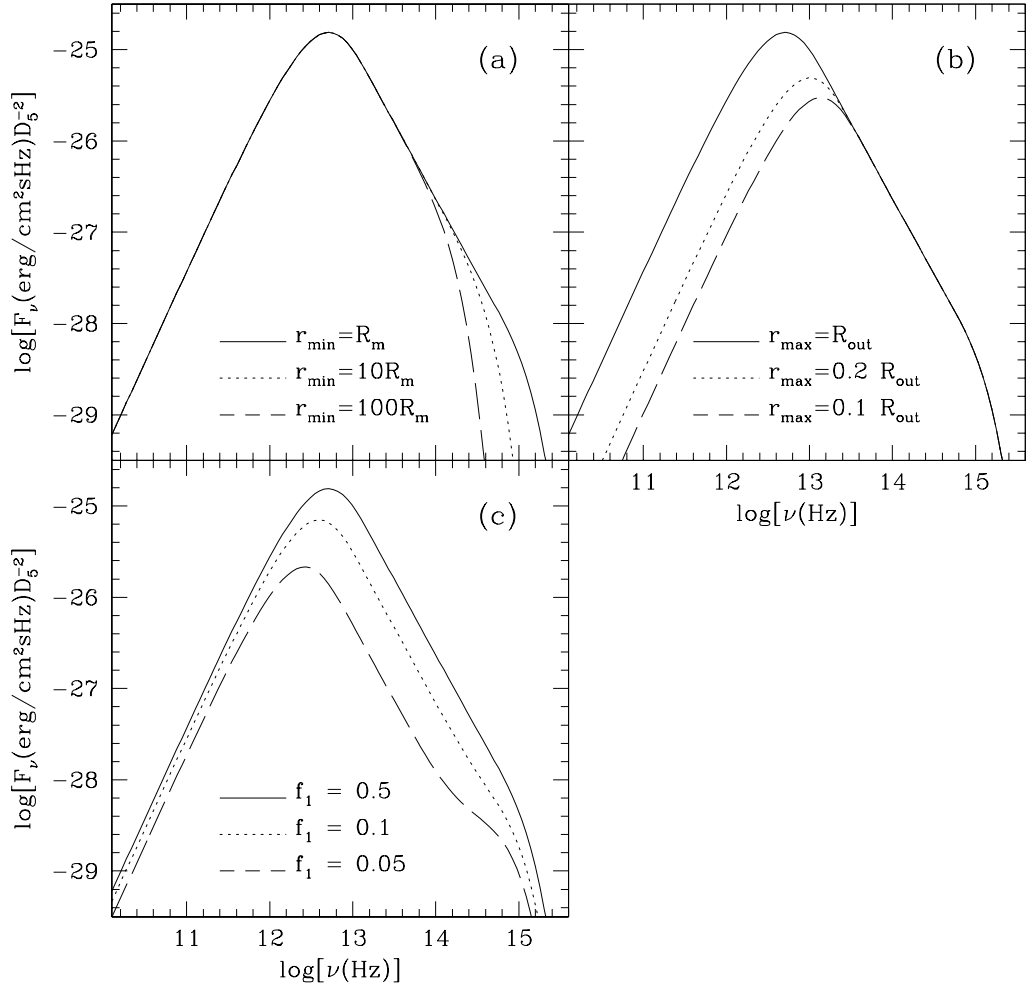}}
\caption{The effect on the
emitted flux of reducing the inner [panel (a)] or outer 
[panel (b)] edge of the disk, or having a larger albedo [panel (c)] 
 is shown for $L^0_x=10^{35}$ ergs sec$^{-1}$ and isotropic
$X$-ray emission.}
\label{fig:4}
\end{figure}

\end{document}